\documentclass[epj,referee,fleqn]{svjour}

\usepackage{latexsym}
\usepackage{graphics}
\usepackage{amsmath}
\usepackage{amsfonts}
\usepackage{amssymb}
\usepackage[dvips]{color}

\begin{document}
\title{Electric and magnetic form factors of strange baryons}

\author{Tim Van Cauteren\inst{1} \and Dirk Merten\inst{2} \and Tamara
Corthals\inst{1} \and Stijn Janssen\inst{1} \and Bernard
Metsch\inst{2} \and Herbert-R. Petry\inst{2} \and Jan
Ryckebusch\inst{1}}

\institute{Ghent University, Dept. of Subatomic and Radiation
Physics, Proeftuinstraat 86, B-9000 Gent, Belgium \\
\email{Tim.VanCauteren@UGent.be} \and Helmholtz-Institut f\"ur
Strahlen- und Kernphysik, Nu\ss allee 14-16, D-53115 Bonn, Germany}

\date{\today}

\abstract{
Predictions for the electromagnetic form factors of the $\Lambda$,
$\Sigma$ and $\Xi$ hyperons are presented.  The numerical
calculations are performed within the framework of the fully
relativistic constituent-quark model developed by the Bonn group.  The
computed magnetic moments compare favorably with the experimentally
known values. Most magnetic form factors $G_M (Q^2)$ can be
parametrized in terms of a dipole with cutoff masses ranging from $0.79$ to
$1.14$ GeV.
}

\maketitle

\def\Id{{\rm 1\kern-.3em I}}

\section{Introduction}\label{sec:intro_gen}

Ever since the pioneering work by Murray Gell-Mann
\cite{gellmann1,gellmann2}, Yuval Ne'eman~\cite{neeman} and George
Zweig~\cite{zweig1,zweig2}, the concept of \emph{constituent quarks}
(CQ) has become well accepted in hadronic physics. To date,
constituent quarks are the effective degrees of freedom in many
existing models for hadrons. They represent, however, not the
fundamental degrees of freedom of the theory of strong interactions,
quantum chromodynamics (QCD). Usually, one connects the effective and
fundamental degrees of freedom by noting that constituent quarks are
conglomerates of quarks, antiquarks and gluons, such that the quantum
numbers of the composed hadron depend only on those of the
conglomerate. Thereby, one presumes that the quark and gluon
degrees of freedom can be efficiently described by means of
constituent quarks in the energy domain where the full QCD equations
cannot be solved perturbatively. This procedure results in equations
that are admittedly easier to handle than those obtained within the
framework of nonperturbative QCD, but still carry all the complications
connected with the (relativistic) treatment of a two- and three-body
problem. The major justification for CQ models is their effectiveness
in describing hadron spectra, symmetry properties and electromagnetic
form factors. This work will focus on the latter quantities.

Meson photoproduction and scattering are primary tools to gain a
deeper insight into the dynamics of baryon resonances. Fair
descriptions of these data can be reached within the framework of
isobar models.  These models typically adopt hadrons and their
resonances as effective degrees of freedom.  Their finite size is
modeled through the introduction of hadronic and electromagnetic form
factors.  With the eye on optimizing the agreement between
calculations and data, the cutoff masses entering the form factors and
the coupling constants are often treated as parameters.  The
underlying philosophy is that the fitted values can subsequently be
compared to predictions from more fundamental models which explicitly
account for the hadron substructure and dynamics.  On the other hand,
the coupling constants, computed in CQ models, could serve as
input parameters into isobar descriptions of meson photoproduction
processes, thereby establishing a more direct link between models for
hadron structure and meson-photoproduction and -scattering
experiments.

Attempts to describe meson-production processes within the framework of
a CQ model include the following ones. Zhao and Li \emph{et al.} have
used a (chiral) CQ model for baryons and quark-meson couplings to
describe \emph{e.g.} $\omega$ and $\eta$ photoproduction on the
proton~\cite{zhao01,li98}. Oh \emph{et al.}~\cite{oh99} have
investigated the contributions of direct knockout, diffractive and
one-boson-exchange processes in $\phi$ electroproduction. In
Ref.~\cite{kroll97}, a diquark-quark model has been used to
calculate kaon-photoproduction cross sections. A dynamical approach to
predict $\pi N$ scat\-te\-ring amplitudes has been developed in
Refs.~\cite{sato96,yoshimoto00}.

The work presented here finds its motivation in the development of a
consistent description of kaon-production processes $p(\gamma,K^+)Y$
and $p(e,e'K^+)Y$~\cite{stijn1,stijn2,stijn3,stijn4}, based on CQ
degrees of freedom. New data for these processes have recently been
released by the CLAS collaboration at Jefferson
Laboratory~\cite{CLAS03} and by the LEPS collaboration at
SPring-8~\cite{zegers03}. Also the SAPHIR collaboration at ELSA in
Bonn~\cite{saphir2003} and the GRAAL collaboration at
Grenoble~\cite{graal2001} will provide extensive data sets for kaon
photoproduction in the very near future. The abundant amount of new
data calls for an appropriate theoretical treatment covering the
complete data base.  One of the major sources of theoretical
un\-cer\-tain\-ties when modeling $p(e,e'K)Y$ reactions, is the
$Q^2$ de\-pen\-dence of the electromagnetic form factors of the
``strange'' baryons~\cite{stijn4}.  In this work, theoretical
predictions for these quantities will be presented.

In the resonance region, pion and eta photoproduction on the proton
can be reasonably described within the framework of isobar models. For
the pion channel, this success can be mainly attributed to the
dominant role of the $\Delta (1232)$ resonance in the reaction
dynamics. The large coupling of this resonance to the $\pi N$ decay
channel makes contributions of other reaction mechanisms seem like
rather small perturbations. A similar role is played by the
$S_{11}(1535)$ resonance in $\eta$ photo- and electroproduction. In
comparison to $\pi$ and $\eta$ production, kaon photo- and
electroproduction is more difficult to treat, since there is no
obvious dominant reaction mechanism, but several contributions
compete. Furthermore, the threshold for production of strange
$s\bar{s}$ pairs increases the energy scale to a domain in which
isobar models could be expected to start losing their validity. A CQ
model could provide an alternative approach, since its number of free
parameters remains low, no matter how many resonances participate in
the production mechanism. Also, CQ's are supposed to be smaller in
size than the hadron they represent~\cite{petronzio03}. Therefore, CQ
models are expected to be valid up to larger energies and momentum
transfers.

Many CQ approaches start off nonrelativistically and require
relativistic corrections at some point. This procedure of relativizing
certain aspects of the model usually involves some degree of
arbitrariness. The CQ model which will be applied here, on the other
hand, has been developed by the Bonn
group~\cite{loering1,loering2,loering3,mertenphd} and is
relativistically covariant in its inception. Yet, at the same time, it
is linked to nonrelativistic models in a transparent way.  The latter
feature arises from the use of the instantaneous approximation and the
$CPT$ theorem which ensures that we arrive at the same number of bound
states as nonrelativistic models~\cite{loering1}. In addition an
extended version of a harmonic-oscillator basis, which also serves as
the starting point of many nonrelativistic CQ models is used.  The
Bonn CQ model is primarily based on the Bethe-Salpeter
approach~\cite{salpeter}. The quantities of physical interest can be
obtained from integral equations which are solved
numerically. Thereby, some freedom exists with respect to the
plausible types of interactions between the constituent quarks. In
order to preserve Lorentz covariance, which is mandatory for
describing boosts consistently, it is assumed that the inter-quark
forces do not depend on the components of the variables parallel to
the total four-momentum of the baryon. In the rest frame, this means
that the interactions are instantaneous or, in other words,
independent of the energy components of the variables. This has the
numerical advantage of reducing the integrations from eight to six
dimensions when determining the Bethe-Salpeter amplitudes. The
relativistic CQ model developed in Bonn adopts a typical linear
confinement potential ($V_c$) supplemented by the 't Hooft instanton
induced interaction ($V_{III}$). This approach allows one to use
merely seven free parameters.  They can be constrained by means of the
mass spectra of strange and nonstrange
baryons~\cite{loering2,loering3}.

Previous work on electromagnetic form factors using the Bethe-Salpeter
approach has been reported in Refs.~\cite{koll00} for mesons and in
Ref.~\cite{merten1} for nonstrange baryons. An excellent description
of the lowest pseu\-dosca\-lar- and vector-meson elastic and
transition form factors was obtained, except for the pion isotriplet,
where the outcome was reasonable. The results on electromagnetic
properties of the nonstrange ba\-ry\-ons and baryon resonances are in
quantitative agreement with the existing data up to the third
resonance region ($W \leq 1.7$ GeV). It should be stressed that in
many investigations, the covariant description of the dynamics turned
out to be of the utmost importance~\cite{wagenbrunn01}.

This work focuses on computing the electric and magnetic form factors
of strange baryons, as well as the electromagnetic form factors of the
$\Sigma^0 \rightarrow \Lambda$ transition. In
Sec.~\ref{sec:formalism}, the Bethe-Salpeter (BS) formalism will be
sketched.  We will then turn our attention to electromagnetic form
factors in Sec.~\ref{sec:formfactors}. The results of our numerical
calculations will be presented in Sec.~\ref{sec:results}.  Whenever
possible we will compare our predictions with experimental data and
previous CQ calculations.

\section{Formalism}\label{sec:formalism}

The Bethe-Salpeter (BS) formalism used in this work is described in
great detail in Refs.~\cite{loering1} and~\cite{merten1}. Here, we
briefly recall its basic ingredients.

\subsection{The Bethe-Salpeter Equation (BSE)}\label{sec:BSE}

In the model adopted here, the basic quantity 
describing a baryon is the three-quark BS amplitude~:
\begin{multline}\label{eq:BSA}
\chi_{\bar{P},a_1,a_2,a_3} (x_1,x_2,x_3) \equiv \\
\langle 0 | T \bigl( \Psi_{a_1}(x_1) \Psi_{a_2}(x_2) \Psi_{a_3}(x_3)
\bigr) |
\bar{P} \rangle \; ,
\end{multline}
where $T$ is the time ordering operator acting on the Heisenberg
quark-field operators $\Psi_{a_i}$ and $\bar{P}$ is the total four-mo\-men\-tum of
the baryon with $\bar{P}_\mu \bar{P}^\mu = M^2$. The $a_i$ denote the
quantum numbers in Dirac, flavor and color space. The
Fourier transform of the above quantity is defined by~:
\begin{multline}
\chi_{\bar{P},a_1,a_2,a_3} (x_1,x_2,x_3) = e^{-i\bar{P}.X}
\chi_{\bar{P}} (\xi,\eta) \\
\equiv e^{-i\bar{P}.X} \int \frac{d^4p_{\xi}}{(2\pi)^4}
\frac{d^4p_{\eta}}{(2\pi)^4} e^{-ip_{\xi}.\xi} e^{-ip_{\eta}.\eta}
\chi_{\bar{P}} (p_{\xi},p_{\eta}) \; ,
\label{eq:BSAF}
\end{multline}
where the scalar product of two four-vectors is given by the
convention $a.b = a_\mu b^\mu = a_0 b^0 - \vec{a}.\vec{b}$. The
standard definition of the Jacobi coordinates and momenta is adopted~:
\begin{subequations}
\label{eq:jacobi}
\begin{align}
&\left\{ \begin{array}{ccc}
X & = & \frac{1}{3} (x_1 + x_2 + x_3) \; ,\\
\xi & = & x_1 - x_2 \; ,\\
\eta & = & \frac{1}{2} (x_1 + x_2 - 2x_3) \; ,
\end{array} \right.
\label{eq:jacobi_co}\\
&\textrm{and} \nonumber \\
&\left\{ \begin{array}{ccc}
P & = & p_1 + p_2 + p_3 \; ,\\
p_\xi & = & \frac{1}{2} (p_1 - p_2) \; ,\\
p_\eta & = & \frac{1}{3} (p_1 + p_2 - 2p_3) \; .
\end{array} \right.
\label{eq:jacobi_mo}
\end{align}
\textrm{ }
\end{subequations} 
From Eq.~(\ref{eq:BSAF}) it becomes clear that the total momentum
$\bar{P}$ of the baryon enters the definition of the BS amplitude only
parametrically and not as a variable, thereby naturally obeying the
symmetry requirement of translational invariance.

\begin{figure*}
\begin{center}
\resizebox{0.75\textwidth}{!}{\includegraphics{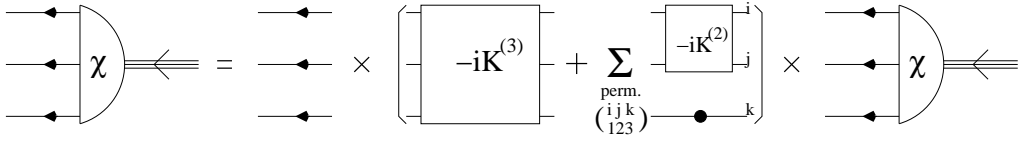}}
\caption{The BS equation in a schematic form. Arrows represent
quark propagators, a filled dot denotes an inverse propagator.}
\label{diag:BSE}
\end{center}
\end{figure*}

The so-called BS amplitude $ \chi_{\bar{P}} \equiv \chi_{\bar{P}}
(p_\xi,p_\eta)$ is the solution to the BS equation \cite{salpeter} which
in momentum space can be schematically written as
\begin{equation}
\chi_{\bar{P}} =
-iG_{0\bar{P}} \left( K^{(3)}_{\bar{P}}+ \bar{K}^{(2)}_{\bar{P}}
\right) \chi_{\bar{P}} \; .
\label{eq:BSE}
\end{equation}
Here, the arguments and integrals over dummy arguments have been
dropped. Its Feynman-diagram analogue is depicted in
Fig.~\ref{diag:BSE}. This equation can be obtained from considering
the six-point Green's function, a quantity which depends on the total
four-momentum squared $P_\mu P^\mu$ and possesses poles at the masses
$M^2$ of the 3-quark bound states. The residue at these poles
corresponds to the product of the BS amplitude and its adjoint.

The quantity $G_{0\bar{P}}$ in Eq.~(\ref{eq:BSE}) is the direct product
of the dressed propagators of the three quarks~:
\begin{multline}
G_{0\bar{P}} (p_{\xi},p_{\eta};p'_{\xi},p'_{\eta}) =
S^1_F\left(\frac{1}{3}P+p_{\xi}+\frac{1}{2}p_{\eta}\right) \\
\textrm{   }\otimes S^2_F\left(\frac{1}{3}P-p_{\xi}+\frac{1}{2}p_{\eta}\right)
\otimes S^3_F\left(\frac{1}{3}P-p_{\eta}\right) \\
\textrm{   }\times (2\pi)^4\delta^{(4)}\left(p_{\xi}-p'_{\xi}\right)
(2\pi)^4\delta^{(4)}\left(p_{\eta}-p'_{\eta}\right) \; .
\label{eq:G0P}
\end{multline}
These propagators are approximated by the propagators of free
constituent quarks. Therefore, we adopt the form
\begin{equation}
S^i_F(p_i) = \frac{i}{\not p_i-m_i+i\epsilon} \; ,
\label{eq:freeprop}
\end{equation}
where $m_i$ is the effective mass of the $i$'th constituent quark.
The quantity denoted by $K^{(3)}_{\bar{P}}$ is the three-particle irreducible
interaction kernel. Further, $\bar{K}^{(2)}_{\bar{P}}$ is a sum of
two-particle irreducible interaction kernels, each multiplied by the inverse
of the propagator of the spectator quark~:
\begin{multline}
\bar{K}^{(2)}_{\bar{P}}
\left(p_{\xi},p_{\eta};p'_{\xi},p'_{\eta}\right) =
K^{(2)}_{(\frac{2}{3}P+p_{\eta})}\left(p_{\xi},p'_{\xi}\right) \\
\otimes \biggl[ S^{3}_F\left(\frac{1}{3}P-p_{\eta}\right) \biggr]^{-1}
\times (2\pi)^4\delta^{(4)}\left(p_{\eta}-p'_{\eta}\right) \\
\textit{   + cycl. perm. in quarks (123)} \; .
\label{eq:barK2}
\end{multline}
In the case of instantaneous forces, $K^{(3)}_{\bar{P}}$ and $K^{(2)}_{p_i+p_j}$
are independent of the component of the Jacobi momenta parallel to the
baryon four-momentum $\bar{P}$, as was already discussed in
Sec.\ref{sec:intro_gen}. In the c.o.m. frame, this condition implies
that there is no dependence on the energy components~:
\begin{subequations}
\label{eq:kernels}
\begin{alignat}{2}
&K^{(3)}_P \left(p_{\xi},p_{\eta};p'_{\xi},p'_{\eta}\right)
\bigg|_{P=(M,\vec{0})} &= &V^{(3)}
\left(\vec{p}_{\xi},\vec{p}_{\eta};
\vec{p'}_{\xi},\vec{p'}_{\eta}\right) \label{eq:kernels3} \; ,\\
&K^{(2)}_{(\frac{2}{3}P+p_{\eta})} \left(p_{\xi},p'_{\xi}\right)
\bigg|_{P=(M,\vec{0})} &= &V^{(2)}
\left(\vec{p}_{\xi},\vec{p}'_{\xi}\right) \label{eq:kernels2} \; .
\end{alignat}
\end{subequations}
We should mention here that whenever a quantity is to be evaluated in the
rest frame of the baryon, we will indicate this by the index $M$, to
make it clear that in this case $\bar{P} = (M,\vec{0})$.

The potentials used in our calculations are those of model
$\mathcal{A}$ of Ref.~\cite{loering2}. The three-particle interaction is
given by a \emph{confinement} potential $V^{(3)}_{conf}$ which rises
linearly with the sum of the distances between the three CQ's. The
two-particle residual interaction is the \emph{'t Hooft Instanton
Induced Interaction} $V^{(2)}_{III}$, which acts between pairs of
quarks that have antisymmetric spin, flavor and color wave functions.

\subsection{Reduction to the Salpeter Equation}\label{sec:reduction}

Solving Eq.~\ref{eq:BSE} can be simplified by exploiting the
instantaneous property of the interaction kernels. Indeed, the
integration over the energy components of the Jacobi momenta can be
performed analytically. This gives rise to a new object $\Phi _M$, the
Salpeter amplitude, which can be directly obtained from the full BS
amplitude~:
\begin{equation}
\Phi_M\left(\vec{p}_{\xi},\vec{p}_{\eta}\right) =
\int \frac{dp^0_{\xi}}{(2\pi)} \frac{dp^0_{\eta}}{(2\pi)}
\chi_M\left(
(p^0_{\xi},\vec{p}_{\xi}),(p^0_{\eta},\vec{p}_{\eta}) \right) \; .
\label{eq:SA}
\end{equation}
This definition is only workable in the special case that no genuine
two-particle irreducible interactions contribute, \emph{e.g.} for the
decuplet baryons which have symmetric spin wave functions. For the
octet baryons, a slightly different approach is needed, as is
explained in the Appendix of Ref.~\cite{merten1} and in
Ref.~\cite{loering1}. There, it is pointed out that for reconstructing
the Bethe-Salpeter amplitude~(\ref{eq:BSA}), it suffices to compute
the projection of the Salpeter amplitude~(\ref{eq:SA}) onto the purely
positive-energy and negative-energy states. This can be accomplished
in the standard manner by introducing the energy projection
operators~:
\begin{equation}
\Lambda^{\pm}_i\left(\vec{p}_i\right) = 
\frac{\omega_i\left(\vec{p}_i\right) \Id \pm H_i\left(\vec{p}_i\right)}
{2\omega_i\left(\vec{p}_i\right)} \; ,
\label{eq:energyprojector}
\end{equation}
where $\omega _i (\vec{p}_i) = \sqrt{m_i^2 + |\vec{p}_i|^2}$ denotes
the energy and
\begin{equation}
H_i(\vec{p}_i)= \gamma^0(\vec{\gamma}.\vec{p}_i+m_i)
\end{equation}
is the free Hamiltonian of the $i$'th quark.  With the above
definition, we define the projected Salpeter amplitude as~:
\begin{multline}
\Phi^{\Lambda}_M\left(\vec{p}_{\xi},\vec{p}_{\eta}\right) =
\left( \Lambda^{+++}\left(\vec{p}_{\xi},\vec{p}_{\eta}\right) +
\Lambda^{---}\left(\vec{p}_{\xi},\vec{p}_{\eta}\right) \right) \\
\times \int \frac{dp^0_{\xi}}{(2\pi)} \frac{dp^0_{\eta}}{(2\pi)}
\chi_M\left( (p^0_{\xi},\vec{p}_{\xi}),
(p^0_{\eta},\vec{p}_{\eta}) \right) \; ,
\label{eq:projSA}
\end{multline}
where $\Lambda^{+++}\left(\vec{p}_{\xi},\vec{p}_{\eta}\right) =
\Lambda^+_1 \left(\vec{p}_1\right) \otimes \Lambda^+_2
\left(\vec{p}_2\right) \otimes \Lambda^+_3
\left(\vec{p}_3\right)$ and
$\Lambda^{---}\left(\vec{p}_{\xi},\vec{p}_{\eta}\right) =
\Lambda^-_1 \left(\vec{p}_1\right) \otimes \Lambda^-_2
\left(\vec{p}_2\right) \otimes \Lambda^-_3
\left(\vec{p}_3\right)$.

\begin{figure*}
\begin{center}
\resizebox{0.75\textwidth}{!}{\includegraphics{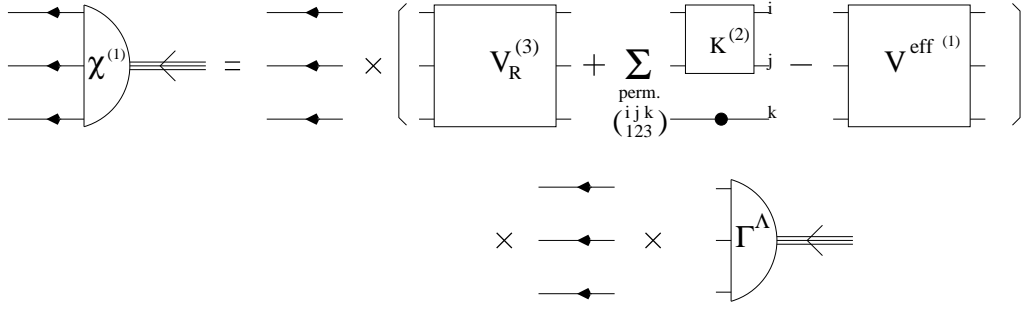}}
\caption{The reconstruction of the BS amplitude from the vertex
function according to Eq.~(\ref{eq:approxBS}).}
\label{diag:BSE2}
\end{center}
\end{figure*}

The Salpeter equation is now given by~:
\begin{multline}
\Phi^{\Lambda}_M \left( \vec{p}_{\xi}, \vec{p}_{\eta} \right)
= \biggl[ \frac{\Lambda^{+++} \left( \vec{p}_{\xi}, \vec{p}_{\eta}
\right)} {M - \Omega \left( \vec{p}_{\xi}, \vec{p}_{\eta}
\right) + i\varepsilon} \phantom{\biggr]} \\
\phantom{\biggl[} + \frac{\Lambda^{---} \left( \vec{p}_{\xi},
\vec{p}_{\eta} \right)} {M + \Omega \left( \vec{p}_{\xi},
\vec{p}_{\eta} \right) - i\varepsilon} \biggr] 
\gamma^0 \otimes \gamma^0 \otimes \gamma^0 \\
\times \int
\frac{d^3p'_\xi} {(2\pi)^3} \frac{d^3p'_\eta} {(2\pi)^3}
V^{(3)} \left( \vec{p}_{\xi}, \vec{p}_{\eta}; \vec{p}'_{\xi},
\vec{p}'_{\eta} \right) \Phi^{\Lambda}_M \left( \vec{p}'_{\xi},
\vec{p}'_{\eta} \right) \\
+ \left[ \frac{\Lambda^{+++} \left( \vec{p}_{\xi}, \vec{p}_{\eta}
\right)} {M - \Omega \left( \vec{p}_{\xi}, \vec{p}_{\eta}
\right) + i\varepsilon} - \frac{\Lambda^{---} \left( \vec{p}_{\xi},
\vec{p}_{\eta} \right)} {M + \Omega \left( \vec{p}_{\xi},
\vec{p}_{\eta} \right) - i\varepsilon} \right] \\
\times \int \frac{d^3p'_\xi} {(2\pi)^3} \biggl[ \left[ \gamma^0 \otimes
\gamma^0 V^{(2)} \left( \vec{p}_{\xi}, \vec{p}'_{\xi} \right)
\right] \otimes \Id \; \biggr] \Phi^{\Lambda}_M \left( \vec{p}'_{\xi},
\vec{p}_{\eta} \right) \\
+ \textit{ cycl. perm. in quarks (123)} \; ,
\label{eq:salpeq}
\end{multline}
where $\Omega \left( \vec{p}_{\xi}, \vec{p}_{\eta} \right)$ is
the sum of the energies of the three constituent quarks~:
\begin{equation}
\Omega = \sum ^{3} _{i=1} \omega_i = \sum ^{3} _{i=1}
\sqrt{|\vec{p}_i|^2+m^2_i} \; .
\label{eq:kin_energy}
\end{equation}
Once the Salpeter Eq.~(\ref{eq:salpeq}) is solved,
the vertex function $\Gamma^\Lambda_M$ can be constructed~:
\begin{multline}
\Gamma^\Lambda_M \left( \vec{p}_\xi, \vec{p}_\eta \right) = -i
\int \frac{d^3p'_\xi} {(2\pi)^3} \frac{d^3p'_\eta} {(2\pi)^3}
\biggl[ V^{(3)}_\Lambda \left( \vec{p}_\xi, \vec{p}_\eta;
\vec{p}'_\xi, \vec{p}'_\eta \right) \phantom{\biggr]} \\
\phantom{\biggl]} + V^{eff^{(1)}}_M \left( \vec{p}_\xi, \vec{p}_\eta;
\vec{p}'_\xi, \vec{p}'_\eta \right) \biggr] \Phi^{\Lambda,(1)}_M
\left( \vec{p}'_\xi, \vec{p}'_\eta \right) \; .
\label{eq:vertexfunc}
\end{multline}
This vertex function describes how the three CQ's couple to form a
baryon, and in first order can be related to the BS amplitude
through~:
\begin{equation}
\chi_{\bar{P}} \approx \chi^{(1)}_{\bar{P}} = \left[ G_{0 \bar{P}} \left( V^{(3)}_R +
\bar{K}^{(2)}_{\bar{P}} - V^{eff^{(1)}}_{\bar{P}} \right) G_{0
\bar{P}} \right] \Gamma^\Lambda_{\bar{P}} \; ,
\label{eq:approxBS}
\end{equation}
of which a diagram is shown in Fig.~\ref{diag:BSE2}.

In Eqs.~(\ref{eq:vertexfunc}) and~(\ref{eq:approxBS}),
$V^{(3)}_\Lambda = \bar{\Lambda} V^{(3)}_R \Lambda$, where
$\bar{\Lambda} = \gamma^0 \otimes \gamma^0 \otimes \gamma^0 \Lambda
\gamma^0 \otimes \gamma^0 \otimes \gamma^0 $, is that part of the
three-body potential which couples only to purely positive-energy and
negative-energy components of the amplitudes. $V^{(3)}_R = V^{(3)} -
V^{(3)}_\Lambda$ is the remaining part which couples to the
mixed-energy components. $V^{eff^{(1)}}_{\bar{P}}$ is a first-order
approximation of an effective potential with three-body structure
which parameterizes the two-body
interaction~\cite{loering1,merten1}. Further,
$\bar{K}^{(2)}_{\bar{P}}$ is defined in Eqs.~(\ref{eq:barK2})
and~(\ref{eq:kernels2}).

\subsection{Current Matrix Elements}\label{sec:CME}

\begin{figure*}
\begin{center}
\resizebox{0.75\textwidth}{!}{\includegraphics{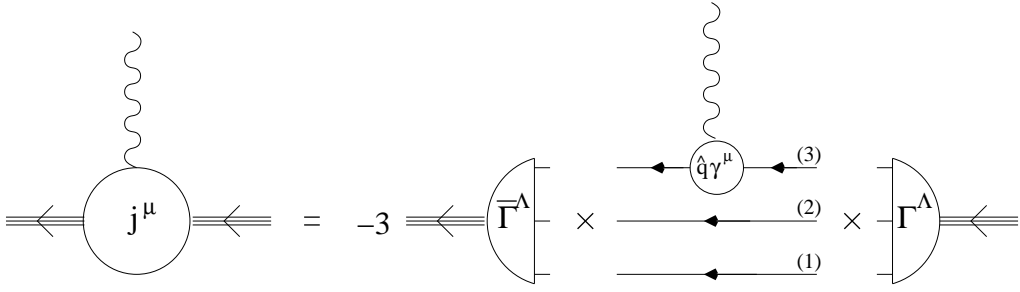}}
\caption{Feynman diagram showing the coupling of the photon to the
third CQ as in Eq.~(\ref{eq:cur_mat_el2}). The other two CQ's are
spectators.}
\label{diag:current}
\end{center}
\end{figure*}

Once the BS amplitudes and vertex functions are determined,  
the current matrix elements can be computed\\
through the following definition 
\begin{equation}
<\bar{P} | j^\mu(x) | \bar{P}'> \quad = \quad <\bar{P} | \bar{\Psi}(x) \hat{q}
\gamma^\mu \Psi(x) | \bar{P}'> \; ,
\label{eq:cur_mat_el1}
\end{equation}
where $\Psi$ and $\hat{q}$ are the constituent-quark field and charge
operator. The above matrix element can be expressed in terms
of the objects defined in Secs.~\ref{sec:BSE}
and~\ref{sec:reduction}. Via
the calculation of six-point and eight-point Green's functions and
residue arguments, it can be shown that in the c.o.m. frame of the
incoming baryon \cite{merten1}
\begin{multline}
<\bar{P} | j^\mu(0) | M> \; \simeq 
-3 \int \frac{d^4p_\xi} {(2\pi)^4}
\frac{d^4p_\eta} {(2\pi)^4} \\
\times \bar{\Gamma}^\Lambda_{\bar{P}} \left(
p_\xi, p_\eta - \frac{2}{3}q \right) S^1_F \left( \frac{1}{3}M + p_\xi
+ \frac{1}{2}p_\eta \right) \\
\otimes S^2_F \left( \frac{1}{3}M - p_\xi + \frac{1}{2}p_\eta \right)
\otimes S^3_F \left( \frac{1}{3}M - p_\eta + q \right) \\
\times \hat{q} \gamma^\mu S^3_F \left( \frac{1}{3}M - p_\eta
\right) \Gamma^\Lambda_M \left( \vec{p}_\xi, \vec{p}_\eta
\right) \; ,
\label{eq:cur_mat_el2}
\end{multline}
where $q$ is the (incoming) photon four-momentum and $\hat{q}$ is the
charge operator working on the third CQ only. Further,
$\bar{\Gamma}^\Lambda_{\bar{P}}$ is the adjoint vertex function and is
calculated in the c.o.m. system according to
\begin{equation}
\bar{\Gamma}^\Lambda_{M} = - \left( \Gamma^{\Lambda_{M}}
\right)^\dagger \gamma^0 \otimes \gamma^0 \otimes \gamma^0 \; .
\label{eq:adj_vert_func}
\end{equation}
Under a Lorentz boost, the vertex function transforms
as~\cite{mertenphd}
\begin{multline}
\Gamma_{\bar{P}} \left(p_\xi,p_\eta - \frac{2}{3} q\right) = \\
S^1_\Lambda \otimes S^2_\Lambda \otimes S^3_\Lambda
\Gamma_{\Lambda^{-1} \bar{P}} \Big(\Lambda^{-1} p_\xi,\Lambda^{-1}
\big(p_\eta - \frac{2}{3} q\big) \Big) \; ,
\label{eq:vert_func_boost}
\end{multline}
with $\Lambda$ the boost matrix and $S^i_\Lambda$ the corresponding
boost operator acting on the $i$'th quark.  Eq.~(\ref{eq:cur_mat_el2})
is a consistent lowest-order approximation of the current matrix
element. We refer to Refs.~\cite{mertenphd} and~\cite{merten1} for
more details and to Fig.~\ref{diag:current} for a schematic
representation of Eq.~(\ref{eq:cur_mat_el2}).  The integration over
the energy variables can be performed analytically.  In the remaining
integral over $\vec{p}_\xi$ and $\vec{p}_\eta $, the azimuthal
dependence can be reduced to $(\phi_\xi - \phi_\eta)$, leaving one
with five-dimensional integrals, which are computed numerically.

\section{Form Factors}\label{sec:formfactors}

In Sec.~\ref{sec:results}, results for the elastic and transition
electromagnetic form factors of the octet baryons with a nonvanishing
strangeness quantum number will be presented. Here, we briefly discuss
our conventions regarding the connection between form factors and
current matrix elements.

\subsection{Elastic Form Factors}\label{sec:el_ff}

The strength with which real and virtual photons couple to baryons can
be quantified in different ways. For the elastic processes, where the
incoming and outgoing baryon are identical, we will compute the Sachs form
factors. We define the vertex function $\Gamma^\mu$ between a baryon
and a photon as~:
\begin{multline}
\langle B, \bar{P}', \lambda' | j^\mu (0) | B, \bar{P}, \lambda
\rangle = e \bar{u}_{\lambda'} (\bar{P}') \Gamma^\mu u_{\lambda}
(\bar{P}) \\
= e \bar{u}_{\lambda'} (\bar{P}') \biggl[ \gamma^\mu F^B_1
(Q^2) + \frac{i \sigma^{\mu\nu} q_\nu} {2M} F^B_2 (Q^2) \biggr] u_{\lambda}
(\bar{P}) \; ,
\label{eq:EMvertex}
\end{multline}
where $B$ denotes the baryon under investigation, $\lambda^{(')}$ the
baryon helicity, $\bar{P}^{(')}$ the baryon on-shell four-momentum and
$u_{\lambda} (\bar{P})$ a Dirac spinor, normalized according to
\begin{equation}
\bar{u}_{\lambda'} (\bar{P}) u_{\lambda} (\bar{P}) =
2M\delta_{\lambda \lambda'} \; .
\end{equation}
The functions $F^B_1$ and $F^B_2$ are
the Dirac and Pauli form factors and depend only on $Q^2 = -q^2$,
where $q$ is the four-momentum carried by the photon. The Sachs form
factors are defined in the standard fashion
\begin{subequations}
\label{eq:sachs_ff}
\begin{align}
G^B_E (Q^2) &= F^B_1 (Q^2) - \frac{Q^2}{4M^2} F^B_2 (Q^2) \; ;
\label{eq:sachs_el} \\
G^B_M (Q^2) &= F^B_1 (Q^2) + F^B_2 (Q^2) \; . \label{eq:sachs_ma}
\end{align}
\end{subequations}

The equations connecting the Sachs form factors to the current matrix
elements in the rest frame of the incoming baryon read~:
\begin{subequations}
\label{eq:S_CME}
\begin{align}
G^B_E (Q^2) &= \frac {\langle B, \bar{P}', \frac{1}{2} | j_0 (0) |
B, \bar{M}, \frac{1}{2} \rangle} {\sqrt{4M^2 + Q^2}} \; ; \label{eq:S_el} \\
G^B_M (Q^2) &= \frac {\langle B, \bar{P}', \frac{1}{2} | j_+ (0) |
B, \bar{M}, -\frac{1}{2} \rangle} {2\sqrt{Q^2}} \; . \label{eq:S_ma}
\end{align}
\end{subequations}

Measurements of the magnetic moments for the strange baryons represent
a direct test of the calculations which will be presented here. These
values should be compared to the values of the magnetic Sachs form
factors at $Q^2 = 0$. From the slope of the form factors at $Q^2 = 0$,
the electric and magnetic mean square radii of the baryons can be
deduced from~:
\begin{equation}
\langle r^2 \rangle = -6 \frac{1}{G(0)} \frac{d G(Q^2)}{d Q^2}
\bigg|_{Q^2=0} \; ,
\label{eq:msrad1}
\end{equation}
if the form factor does not go to zero for $Q^2 \rightarrow 0$, and~:
\begin{equation}
\langle r^2 \rangle = -6 \frac{d G(Q^2)}{d Q^2} \bigg|_{Q^2=0} \; ,
\label{eq:msrad2}
\end{equation}
if the form factor vanishes at $Q^2 = 0$. Two recent measurements at
CERN~\cite{adamovich99} and Fermilab~\cite{eschrich02} provided the
first values for the electric mean square radius of the
$\Sigma^-$  hyperon. To our knowledge, the $\Sigma^-$ is the only hyperon
for which such information is presently available.

\subsection{Transition Form Factors}\label{sec:tr_ff}

When describing electromagnetic transitions at vertex le\-vel, at a
certain point, one is forced to make a choice as to what operatorial
form to use. The only condition which should be obeyed is the Ward
identity $q_\mu \Gamma^\mu = 0$. A general form for the vertex
function for spin-$1/2$ baryons corresponding with $\gamma^* + B^*
\longrightarrow B$ transitions, is~:
\begin{multline}
\Gamma^\mu = F^{B^* B}_1 (Q^2) \left( \gamma^\mu + \frac{q^\mu q^\nu}{Q^2}
\gamma_\nu \right) \\
+ \frac{F^{B^* B}_2 (Q^2) \kappa_{B^*B}} {2M_p} i
\sigma^{\mu\nu} q_\nu \; ,
\label{eq:vert_gauge_inv}
\end{multline}
with $M_p$ the proton mass, $F^{B^* B}_1 (Q^2)$ and $F^{B^* B}_2
(Q^2)$ the two transition form factors, belonging to parts of the
vertex that obey the Ward identity individually, and $\kappa_{B^*B}$
the transition magnetic moment in units of the nuclear magneton
$\mu_N$. In the rest frame of the incoming baryon $B^*$, we get the
following equations for the transition form factors~:
\begin{subequations}
\label{eq:trans_ff_inv}
\begin{align}
e F^{B^* B}_1 (Q^2) &= \frac{Q^2} {Q^+ \sqrt{Q^-}} \nonumber \\
& \times \left[ \frac{M + M^*} {|\vec{P}'|} \mathcal{M}_0 -
\frac{1}{2} \mathcal{M}_+ \right] \; ; \label{eq:trans_inv_1}\\
\frac{e \kappa_{B^*B} F^{B^* B}_2 (Q^2)} {2M_p} &= \frac{-1} {Q^+
\sqrt{Q^-}} \nonumber \\
& \times \left[ \frac{Q^2} {|\vec{P}'|} \mathcal{M}_0 + \frac{M +
M^*} {2} \mathcal{M}_+ \right] \; , \label{eq:trans_inv_2}
\end{align}
\end{subequations}
with $M^*$ and $M$ the mass of incoming and outgoing baryon respectively,
$|\vec{P}'|$ the magnitude of the three-momentum of the outgoing baryon,
$Q^\pm = Q^2 + (M^* \pm M)^2$ and~:
\begin{subequations}
\label{eq:mat_el}
\begin{align}
\mathcal{M}_0 &= \langle B, \bar{P}', \frac{1}{2} | j_0 (0) | B^*,
\bar{M}^*, \frac{1}{2} \rangle \; ; \label{eq:mat_el_0} \\
\mathcal{M}_+ &= \langle B, \bar{P}', \frac{1}{2} | \left(- j^1 (0) -
i.j^2 (0) \right) | B^*, \bar{M}^*, -\frac{1}{2} \rangle \; .
\label{eq:mat_el_+}
\end{align}
\end{subequations}
Hereby, we have implicitly adopted the axial gauge $\vec{\epsilon}
. \vec{q} = 0$, where $\vec{\epsilon}$ is the photon-polarization
three-vector.

With these definitions for the transition form factors, $F^{B^* B}_1
(0)$ gives the \emph{transition} charge and $\kappa_{B^*B}$ is the
transition magnetic moment, since $F^{B^* B}_2 (0) = 1$ by convention.

\section{Results}\label{sec:results}

In this section, results for the computed electric and magnetic form
factors of the strange particles belonging to the baryon octet will be
presented. We will discuss the elastic and the $\Sigma^0 \rightarrow
\Lambda$ transition form factors. Comparisons with other calculations
will be made. In Ref.~\cite{kim96}, Kim \emph{et al.}  present
calculations for the elastic form factors of the ground-state octet
baryons up to $Q^2 = 1.0$ GeV$^2$ within the framework of the
chiral quark/soliton model. Kubis \emph{et al.} have computed electric
and magnetic form factors of the hyperons for $Q^2 < 0.2$ GeV$^2$ in
the framework of heavy-baryon chiral perturbation theory (HB) in
Ref.~\cite{kubis99} and later extended their model to fourth
order~\cite{kubis01} to recalculate the electric form factors of the
baryon octet and the $\Sigma^0 \rightarrow \Lambda$ transition form
factor $F^{B^*B}_1(Q^2)$ for $Q^2 < 0.3$ GeV$^2$. In the same article,
relativistic baryon chiral perturbation employing infrared regulators
(IR) is used and shown to have predictive value. Since these
investigations are confined to small values for $Q^2$, we will only
compare our results for the magnetic moments and the mean square radii
for the HB and IR results. We will also confront our predictions with
those presented in Ref.~\cite{plessas03}, where results are shown of
CQ calculations based on a Goldstone-Boson Exchange (GBE) quark-quark
interaction~\cite{glozman98a,glozman98b} and a One-Gluon Exchange
(OGE) interaction~\cite{bhaduri81,theussl01}.

\begin{figure*}
\begin{center}
\resizebox{0.75\textwidth}{!}{\includegraphics{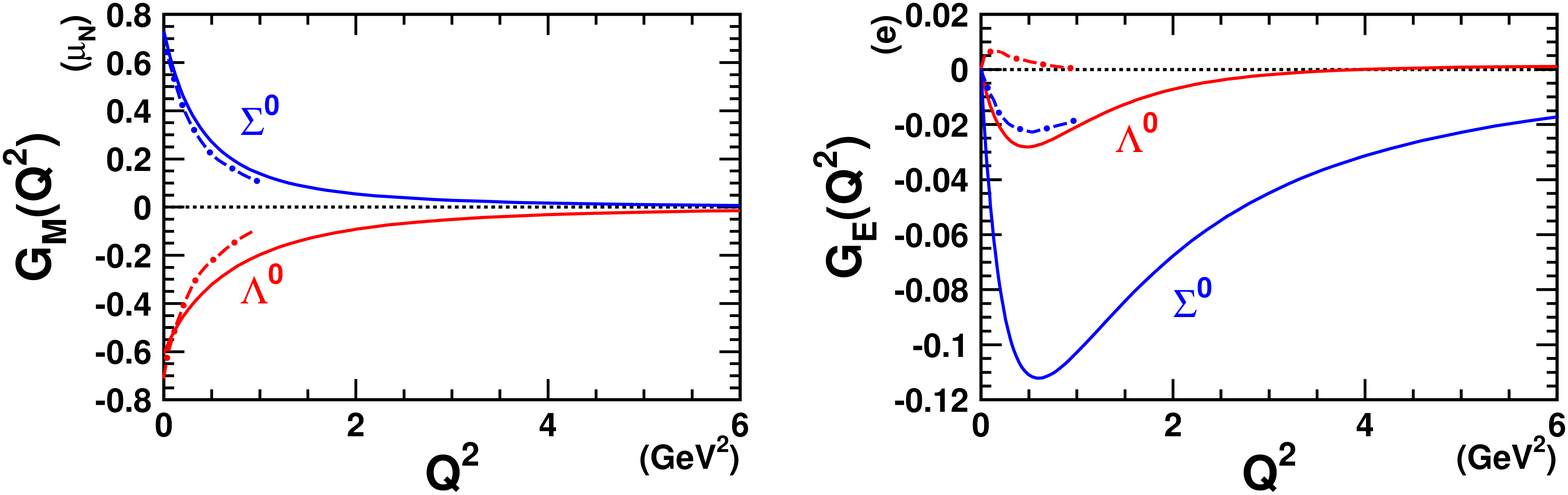}}
\caption{Calculated magnetic (left) and electric (right) form
factors of the $\Lambda$ and $\Sigma^0$ hyperon. The dot-dashed
curves are the predictions from Ref.~\cite{kim96}.}
\label{fig:neut_1}
\end{center}
\end{figure*}

In Fig.~\ref{fig:neut_1}, our results for the neutral single-strange
baryons are displayed. The computed $Q^2$ dependence of the magnetic
form factor of the $\Sigma^0$ and $\Lambda$ hyperon nicely follows
that of a dipole
\begin{equation}
G(Q^2) = \frac{G(0)} {\left( 1 + \frac{Q^2}{\Lambda^2} \right)^2} \; ,
\label{eq:dipole}
\end{equation}
with cutoff masses $\Lambda_M = 0.88$ GeV and $1.14$ GeV
respectively.  Their values in $Q^2=0$ are the magnetic moments
{$\mu_{\Sigma^0} = 0.73$} and $\mu_{\Lambda} = -0.61$ in units of the
nuclear magneton $\mu_N$, which are very
realistic~(Table~\ref{tab:mag_mom}). The electric form factors in the
right panel of Fig.~\ref{fig:neut_1} have the opposite sign in
comparison with the neutron electric form factor. This can be
attributed to the heavier $s$ quark in the hyperons, which has a
higher probability of residing near the center of mass of the hyperon,
making the electric density negative at small $r$, whereas it is
positive for the neutron~\cite{kelly02}. The predicted negative values
for $G_E$ for the $\Sigma^0$ and $\Lambda$, are in contradiction
with the results from Refs.~\cite{kim96} and~\cite{kubis01}. Kim
\emph{et al.} predict a positive $G_E$ for the $\Lambda$ and Kubis
\emph{et al.}  predict a negative mean square radius for the
$\Sigma^0$ hyperon~(Table~\ref{tab:elec_rad}). It should also be noted
that for neutral hyperons, our results for the electric form factors
are about a factor of five larger in magnitude than those of
Ref.~\cite{kim96}. This suggests that in our model, there is a higher
charge density near the center of mass of the neutral hyperon than in
the chiral quark/soliton model.

\begin{figure*}
\begin{center}
\resizebox{0.75\textwidth}{!}{\includegraphics{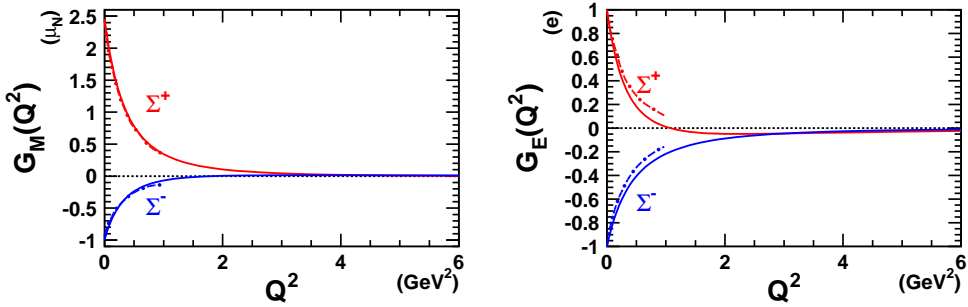}}
\caption{Calculated magnetic and electric form factors of the
$\Sigma^+$ and $\Sigma^-$ hyperon. The dot-dashed curves are the
predictions from Ref.~\cite{kim96}.}
\label{fig:charge_1}
\end{center}
\end{figure*}

Our predictions for the charged single-strange baryons $\Sigma^\pm$
are shown in Fig.~\ref{fig:charge_1}. Again, the results for the
magnetic moments $\mu_{\Sigma^+} = 2.47~\mu_N$ and $\mu_{\Sigma^-} =
-0.99~\mu_N$ are in excellent agreement with
experiment~(Table~\ref{tab:mag_mom}). Whilst the magnetic form factor
of the $\Sigma^+$ resembles a dipole with cutoff $\Lambda_M = 0.79$ GeV, the one for the $\Sigma^-$ drops relatively fast and even
changes sign at $Q^2 \approx 1.6$ GeV, remaining small at high
$Q^2$. A similar qualitative behavior is observed for the electric
form factor of the $\Sigma^+$, changing sign at $Q^2 \approx 1.1$
GeV$^2$. For $Q^2 > 2.6$ GeV$^2$, the form factors of $\Sigma^+$ and
$\Sigma^-$ become practically indistinguishable. Inspecting
Fig.~\ref{fig:charge_1}, it is clear that our predictions for the
magnetic form factors agree remarkably well with those of
the chiral quark/soliton model at low values of $Q^2$~\cite{kim96}.

To our knowledge, for the electric mean square radius of the
$\Sigma^-$ hyperon, the following experimental values are presently
available~:
\begin{equation}
<r_E^2>_{\Sigma^-} = 0.60 \pm 0.08 \textrm{ (stat.) } \pm 0.08 \textrm{
(syst.) fm}^2
\label{eq:r_E_FNAL}
\end{equation}
from Ref~\cite{eschrich02}, and~:
\begin{equation}
<r_E^2>_{\Sigma^-} = 0.91 \pm 0.32 \textrm{ (stat.) } \pm 0.40 \textrm{
(syst.) fm}^2
\label{eq:r_E_CERN}
\end{equation}
from Ref~\cite{adamovich99}. Our prediction $<r_E^2>_{\Sigma^-} =
0.49$ fm$^2$ (Table~\ref{tab:elec_rad}) is compatible with both these
values.

\begin{figure*}
\begin{center}
\resizebox{0.75\textwidth}{!}{\includegraphics{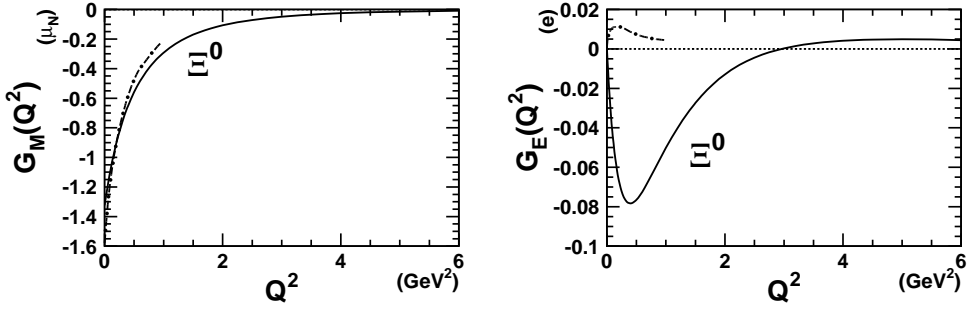}}
\caption{Calculated magnetic and electric form factors of the $\Xi^0$
hyperon. The dot-dashed curves are the predictions from Ref.~\cite{kim96}.}
\label{fig:xi_n}
\end{center}
\end{figure*}

\begin{figure*}
\begin{center}
\resizebox{0.75\textwidth}{!}{\includegraphics{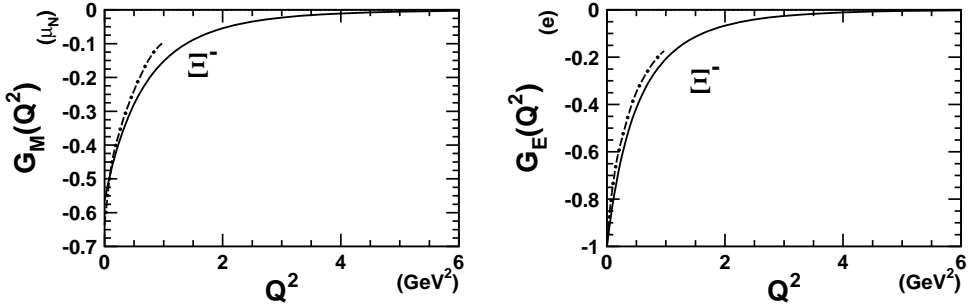}}
\caption{Calculated magnetic and electric form factor of the
$\Xi^-$ hyperon. The dot-dashed curves are the predictions from
Ref.~\cite{kim96}.}
\label{fig:xi_m}
\end{center}
\end{figure*}

The experimental information regarding the $\Xi$ doublet is scarce. To
complete the description of ground-state hyperons, we have calculated
its elastic form factors. The form factors of the $\Xi^0$ are
displayed in Fig.~\ref{fig:xi_n}.  The $G_E(Q^2)$ changes sign about
$Q^2 = 3.0$ GeV$^2$ and $G_M(Q^2)$ can be nicely fitted with a dipole
with $\Lambda_M = 0.94$ GeV and magnetic moment $\mu_{\Xi^0} =
-1.33~\mu_N$. Again, this value for $\mu_{\Xi^0}$ is in good agreement
with the experimentally determined value
(Table~\ref{tab:mag_mom}). The $\Xi^-$ exhibits dipole-like behavior
in both $G_E(Q^2)$ and $G_M(Q^2)$ (Fig.~\ref{fig:xi_m}) with cutoffs
$\Lambda_E = 0.93$ GeV and $\Lambda_M = 1.03$ GeV, respectively. Our
prediction for the magnetic moment, $\mu_{\Xi^-} = -0.57~\mu_N$, is
close to the experimental value $-0.6507 \pm
0.0025~\mu_N$~\cite{PDG2002}.

As is the case for the $\Lambda$ hyperon, the calculations of
Ref.~\cite{kim96} predict a different sign and a smaller magnitude for
the electric form factor of the neutral state of the $\Xi$ doublet,
but the other results agree very well.

\begin{table*}
\begin{center}
\caption{Magnetic moments of strange baryons in units of $\mu_N$.  The
notation \emph{GBE/OGE} (\emph{HB/IR}) refers to the two different
models discussed in Ref.~\cite{plessas03} (\cite{kubis01}). In
Ref.~\cite{kubis01}, only the transition magnetic moment for $\Sigma^0
\rightarrow \Lambda$ is a real prediction. Experimental values are
taken from Ref.~\cite{PDG2002}, except for $\mu_{\Sigma^0} =
(\mu_{\Sigma^+} + \mu_{\Sigma^-}) / 2$, for which isospin invariance
is used. For the $\Sigma^0 \rightarrow \Lambda$ transition, the
absolute value is given.}
\label{tab:mag_mom}
\begin{tabular}{|c|c|c|c|c|c|}
\hline\noalign{\smallskip} 
Baryon & $\mu^{exp}_{Y}$ & $\mu^{calc}_{Y}$ &
$\mu^{\textrm{\cite{kim96}}}_{Y}$ &
$\mu^{\textrm{\cite{plessas03}}}_{Y \: (GBE/OGE)}$ &
$\mu^{\textrm{\cite{kubis01}}}_{Y \: (HB/IR)}$ \\
\noalign{\smallskip}\hline\noalign{\smallskip}
$\Lambda^0(1116)$ & $-0.613 \pm 0.004$ & $-0.61$ & $-0.77$ & $-0.59 /
-0.59$ & exp. \\
$\Sigma^+(1189)$ & $2.458 \pm 0.010$ & $2.47$ & $2.42$ & $2.34 / 2.20$ & exp. \\
$\Sigma^0(1189)$ & $0.649$ & $0.73$ & $0.75$ & $0.70 / 0.66$ & exp. \\
$\Sigma^-(1189)$ & $-1.160 \pm 0.025$ & $-0.99$ & $-0.92$ & $-0.94 / -0.89$ & exp. \\
$| \Sigma^0 \rightarrow \Lambda |$ & $1.61 \pm 0.08$ & $1.41$ & $1.51$
& ---  & $1.46 / 1.61$ \\
$\Xi^0(1315)$ & $-1.250 \pm 0.014$ & $-1.33$ & $-1.64$ & $-1.27 / -1.27$ & exp. \\
$\Xi^-(1315)$ & $-0.6507 \pm 0.0025$ & $-0.57$ & $-0.68$ & $-0.67 / -0.57$ & exp. \\
\noalign{\smallskip}\hline
\end{tabular}
\end{center}
\end{table*}

\begin{table*}
\begin{center}
\caption{Magnetic mean square radii of strange baryons in units of
fm$^2$. All magnetic form factors resemble dipoles, except
for the $\Sigma^-$, and our fitted value for the cutoff mass is
given. The notation \emph{HB/IR} refers to the two models presented in
Ref.~\cite{kubis01}.}
\label{tab:mag_rad}
\begin{tabular}{|c|c|c|c|c|}
\hline\noalign{\smallskip}
Baryon & $<r^2_M>^{calc}$ &
$<r^2_M>^{\textrm{\cite{kim96}}}$ &
$<r^2_M>^{\textrm{\cite{kubis01}}}_{(HB/IR)}$
& $\Lambda_M$ (GeV) \\
\noalign{\smallskip}\hline\noalign{\smallskip}
$\Lambda^0(1116)$ & $0.40$ & $0.70$ & $0.30 / 0.48$ & $1.14$ \\
$\Sigma^+(1189)$ & $0.69$ & $0.71$ & $0.74 /0.80$ & $0.79$ \\
$\Sigma^0(1189)$ & $0.60$ & $0.70$ & $0.20 / 0.45$ & $0.88$ \\
$\Sigma^-(1189)$ & $0.81$ & $0.74$ & $1.33 / 1.20$ & --- \\
$\Sigma^0 \rightarrow \Lambda$ & $1.96$ & --- & $0.60 / 0.72$ & $0.82$ \\
$\Xi^0(1315)$ & $0.47$ & $0.75$ & $0.44 / 0.61$ & $0.94$ \\
$\Xi^-(1315)$ & $0.38$ & $0.51$ & $0.44 / 0.50$ & $1.03$ \\
\noalign{\smallskip}\hline
\end{tabular}
\end{center}
\end{table*}

\begin{table*}
\begin{center}
\caption{Electric mean square radii of strange baryons in units of fm$^2$.
The same conventions as in Tables~\ref{tab:mag_mom} and~\ref{tab:mag_rad}.}
\label{tab:elec_rad}
\begin{tabular}{|c|c|c|c|c|c|c|}
\hline\noalign{\smallskip}
Baryon & $<r^2_E>^{exp}$ & $<r^2_E>^{calc}$ &
$<r^2_E>^{\textrm{\cite{kim96}}}$ &
$<r^2_E>^{\textrm{\cite{kubis01}}}_{(HB/IR)}$
& $<r^2_E>^{\textrm{\cite{plessas03}}}_{(GBE/OGE)}$
& $\Lambda_E$ (GeV) \\
\noalign{\smallskip}\hline\noalign{\smallskip}
$\Lambda^0(1116)$ & --- & $0.038$ & $-0.04$ & $0.00 / 0.11$ & --- &
--- \\
$\Sigma^+(1189)$ & --- & $0.79$ & $0.79$ & $0.72 / 0.60$ & --- & ---
\\
$\Sigma^0(1189)$ & --- & $0.150$ & $0.02$ & $-0.08 / -0.03$ & --- & --- \\
$\Sigma^-(1189)$ & $0.60^{\textrm{\cite{eschrich02}}} /
0.91^{\textrm{\cite{adamovich99}}}$ & $0.49$ & $0.75$ & $0.88 / 0.67$ &
$0.49 / 0.44$ & $0.93$ \\
$\Sigma^0 \rightarrow \Lambda$ & --- & $-0.120$ & --- & $-0.09 / 0.03$ & --- & --- \\
$\Xi^0(1315)$ & --- & $0.140$ & $-0.06$ & $0.08 / 0.13$ & --- & --- \\
$\Xi^-(1315)$ & --- & $0.47$ & $0.72$ & $0.75 / 0.49$ & --- & $0.93$
\\
\noalign{\smallskip}\hline
\end{tabular}
\end{center}
\end{table*}

\begin{figure*}
\begin{center}
\resizebox{0.75\textwidth}{!}{\includegraphics{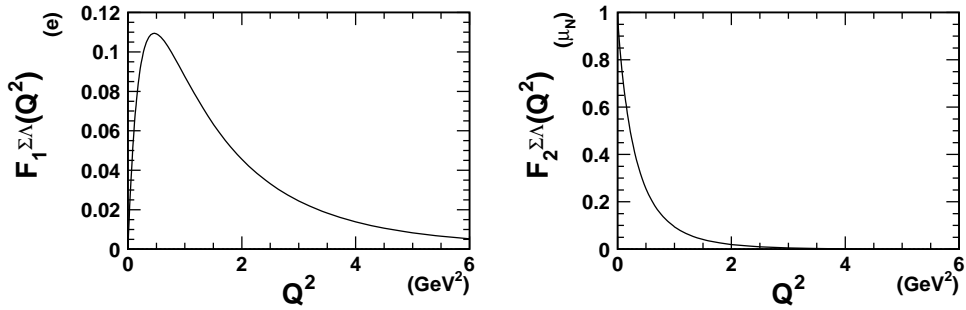}}
\caption{The transition form factors of the $\gamma^* + \Sigma^0
\rightarrow \Lambda$ decay as defined in
Eq.~\ref{eq:trans_ff_inv}.}
\label{fig:siglam}
\end{center}
\end{figure*}

The last point of our discussion concerns the form factors related to
the $\gamma^* + \Sigma^0 \rightarrow \Lambda$ transition. We show the
two form factors $F^{\Sigma \Lambda}_1 (Q^2)$ and $F^{\Sigma
\Lambda}_2 (Q^2)$ in Fig.~\ref{fig:siglam}, calculated with
Eqs.~(\ref{eq:trans_ff_inv}). The only link with experiment is the
transition magnetic moment {$|\mu_{\Sigma \Lambda}| = 1.61 \pm
0.08~\mu_N$} from~\cite{PDG2002}. Our calculated value of
{$|\mu_{\Sigma \Lambda}| = 1.41~\mu_N$} is about $15\%$ off, but still
reasonable considering the relatively large experimental error.

\section{Conclusions}\label{sec:conclusions}

In this work, the first results of an extended implementation of the
Bonn relativistic constituent-quark model into electromagnetic
properties of the strangeness sector have been presented.
Electromagnetic form factors for the low-lying hyperons have been
computed.  Comparison with experimentally determined values is
possible for the magnetic moments and the electric mean square radius
of the $\Sigma^-$ hyperon. A nice agreement between our predictions
and the data is observed.  The predicted $Q^2$ dependence of the form
factors is essential information when modeling kaon electroproduction
processes within an isobar (or, hadrodynamic)
framework~\cite{stijn4}. As illustrated in Sec.~\ref{sec:results}, to
date the different hadron models do not even agree on the sign of the
electric form factors of neutral hyperons.  Some form factors have
been observed to change sign at finite $Q^2$ values.

Work on calculating helicity amplitudes of known and \emph{missing}
hyperon resonances is in progress. We intend to conduct an elaborate
investigation of strong decay widths of baryon resonances into the $K
Y$ channels. In this way we hope to identify, on the basis of
quark-quark dynamics, the most important intermediate baryon
resonances in kaon photo- and electroproduction reactions.


\end{document}